\documentclass[twocolumn,amssymb,pra,superscriptaddress]{revtex4}

\usepackage{graphicx}
\usepackage{enumerate}
\def\mbF{\mathbf{F}}

\begin{document}
\title{A distributed architecture for scalable quantum computation with realistically noisy devices}%

\author{Keisuke Fujii}
\affiliation{Graduate School of Engineering Science, Osaka University,
Toyonaka, Osaka 560-8531, Japan}
\author{Takashi Yamamoto}
\affiliation{Graduate School of Engineering Science, Osaka University,
Toyonaka, Osaka 560-8531, Japan}
\author{Masato Koashi}
\affiliation{Photon Science Center, The University of Tokyo, 2-11-16 Yayoi, 
Bunkyo-ku, Tokyo 113-8656, Japan}
\author{Nobuyuki Imoto}
\affiliation{Graduate School of Engineering Science, Osaka University,
Toyonaka, Osaka 560-8531, Japan}

\date{\today}%

\begin{abstract}
Tremendous efforts have been paid for realization of 
fault-tolerant quantum computation so far.
However, preexisting fault-tolerant schemes assume that 
a lot of qubits live together in a single quantum system,
which is incompatible with actual situations of experiment.
Here we propose a novel architecture for 
practically scalable quantum computation,
where quantum computation is distributed over
small-size (four-qubit) local systems, which are connected by quantum channels. 
We show that the proposed architecture works
even with the error probability 0.1\% of local operations,
which breaks through the consensus
by one order of magnitude.
Furthermore, 
the fidelity of quantum channels
can be very low $\sim$ 0.7,
which substantially relaxes the difficulty of scaling-up the architecture.
All key elements and their accuracy required for the present architecture
are within reach of current technology.
The present architecture allows us to achieve efficient scaling of quantum computer, 
as has been achieved in today's classical computer.
\end{abstract}

\maketitle

\section{Introduction}
Scalability and fault-tolerance are essential ingredients
in both classical and quantum computation.
In 1945, von Neumann proposed the first architecture design
of a classical computer \cite{vonNeumann45},
and most of today's classical computers are of von Neumann type.
He also explored fault-tolerance in classical computation 
by developing a systematic way to construct reliable logic circuits from noisy devices \cite{vonNeumann52},
which is now attracting renewed interest in the field of nanocomputer \cite{Han03}.

In quantum computation, on the other hand,
it is still under extensive investigation 
what type of physical system is best suited to an experimental realization of quantum computation and what level of accuracy is required for it.
For a better understanding of them, a lot of efforts have been made 
over the past decade both theoretically and experimentally.
Quantum fault-tolerance theory ensures scalable quantum computation 
with noisy quantum devices as long as the error probability of such devices
is smaller than a threshold value
(see Ref. \cite{NielsenChuang} and references therein).
The noise thresholds have been calculated to be about $\sim 0.1$--$1\%$
for several fault-tolerant schemes 
under various assumptions
\cite{Knill05,Raussendorf07a,FY10}.
In the experiments, it is nowadays possible to control
a few to dozen qubits in a wide variety of physical systems
such as trapped ions \cite{iontraps} and
nitrogen-vacancy (NV) centres in diamond 
\cite{NVcenter} etc. (see Ref. \cite{review} and references therein).
There is, however, still a large gap between these
top-down (theoretical) and bottom-up (experimental) approaches.
For example, the existing fault-tolerant schemes \cite{Knill05,Raussendorf07a,FY10}
require many qubits live together in a single system 
and assume that the whole system can be controlled
with the same accuracy regardless of its size.
In experiment, on the other hand, 
the number of qubits in a single system
is rather limited; if we increase the number of qubits in a single system, 
the control becomes more and more complex,
which makes it hard to achieve the same accuracy.
In fact,  as mentioned in Ref. \cite{Knill10} there is a consensus that 
``for practical scalability the probability of error introduced 
by the application of a quantum gate must be less than 0.0001".
In order to fill the gap and break through the consensus,
we have to develop a fundamental design of 
a scalable and fault-tolerant architecture for quantum computation 
as von Neumann did in the early years of classical computer technology.

Several architectures for scalable quantum computation
have been proposed so far in bottom-up approaches by
considering specific physical implementations
such as photons \cite{PhotonicQC} and
semiconductor nanophotonics \cite{SemiconductorQC}.
It is, however, still not fully understood what kind of elements
and what level of accuracies of them are essentially
required to construct a scalable and fault-tolerant architecture.
Furthermore, the requirements of the existing quantum architectures
are too demanding and complicated to be experimentally feasible.
In this paper,
we propose a fundamental and simple design of a scalable and fault-tolerant 
architecture for quantum computation taking a top-down approach,
and clarify under what condition scalable quantum computation can be executed fault-tolerantly.

To ensure scalability,
we adopt a distributed approach to quantum computation \cite{Distributed},
where small quantum systems, say {\it quantum arithmetic logic units} (QALUs), are connected via quantum channels.
This type of architectures are equipped with built-in modular scalability. 
That is, the size of the architecture scales up
by adding well-established individual units.
In the previous one-dimensional (1D) distributed architecture \cite{RepeaterQIP,Jian07},
however, entangled states have to be shared between QALUs of 
arbitrary distance, and therefore
the quantum repeater protocol \cite{Repeater} is utilised.
As the distance gets larger, this procedure takes more time.
Moreover, every QALU in-between must provide a workspace for this procedure, resulting in an overhead which grows with the distance.
Besides, if there is a single point of breakdown in the quantum channels 
possibly due to a manufacturing error,
one cannot generate any entanglement between
two parts separated by this point. 

\begin{figure}
\centering
\includegraphics[width=80mm]{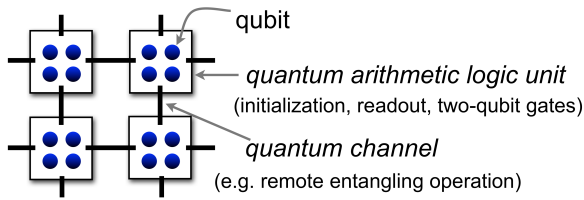}
\caption{
The 2D architecture and QALUs. The 2D distributed architecture,
which consists of four-qubit (blue circles) QALUs (boxes)
connected by quantum channels (lines) with their nearest-neighbours.
}
\label{architecture}
\end{figure}
The present architecture, on the other hand, consists of
a two-dimensional (2D) array of QALUs,
each of which consists of four qubits, as depicted in Fig. \ref{architecture}.
Furthermore, entangled states are 
shared only between the nearest-neighbour QALUs,
which allows highly parallel operations.
Surprisingly, this simple architecture is found to be sufficient 
for fault-tolerant quantum computation
and works with very noisy quantum channels of fidelity $\sim 0.7$
and reasonably accurate local operations of the error probability $\sim 0.1\%$,
which breaks through the consensus by one order of magnitude \cite{Knill10}.
These results are achieved by utilizing twofold error management techniques: 
entanglement purification \cite{Purification1,Purification2,FY09} and topological quantum computation (TQC) \cite{Raussendorf07a}.
The former is employed to implement a reliable two-qubit gate
by using very noisy quantum channels
with the help of quantum gate teleportation \cite{GC99}.
In particular, 
we apply high-performance entanglement purification,
so-called {\it double selection scheme} \cite{FY09},
which is essential for achieving the above result.
TQC is used to handle the remaining errors and 
to archive quantum gate operations of arbitrary accuracy,
which is required for large-scale quantum computation.
Furthermore,
the nearest-neighbour quantum communications together with TQC
is quite robust against the manufacturing error mentioned above,
since we can reconstruct a reliable logical information
on the surface code by avoiding such defects \cite{SemiconductorQC,LossTolerant}.

All key ingredients in the present architecture,
(i) a four-qubit system, (ii) gate operations in the four-qubit system,
and (iii) entangling operations between the separate systems,
have already been demonstrated experimentally in various physical systems.
Actually, the benchmarks in trapped ion systems are 
comparable to the requirements of the proposed architecture.
These results push the realization of large-scale quantum computation 
within reach of current technology.
We believe that this work providing a fundamental and simple design of a scalable and fault-tolerant architecture
for quantum computation fills the gap between the top-down and bottom-up approaches,
and gives a good guideline and benchmark in development of devices for quantum computer.

\section{2D distributed architecture}
The present architecture consists of a 2D array of 
well-defined small (four-qubit) local quantum systems, QALUs (see Fig. \ref{architecture}),
where we can implement the initializations, measurements,
and two-qubit gate operations. 
These local operations are assumed to be imperfect,
which is modeled as follows:
(i) An ideal two-qubit gate is followed by 
two-qubit depolarizing noise,
$(1-p_g)\rho + \sum _{(i,j)\neq (0,0)} p_{ij}
(\sigma _{i}\otimes \sigma _{j}) \rho   (\sigma _{i}\otimes \sigma _{j})$
where $p_g \equiv \sum _{(i,j) \neq (0,0)}  p_{ij} $ ($i,j= 0,1,2,3$),
and $\sigma _{i}$ ($i=0,1,2,3$) are the Pauli matrices ($\sigma _0 =I$).
(ii) The measurement of a physical qubit
is implemented with an error probability $ p_M $.
The QALUs are assumed to work with reasonably high accuracy ($p_g,p_M \sim 0.1\%$).
The memory error probability is assumed to be sufficiently 
smaller than that of the operational errors 
(as is also the case for most of physical systems)
and ignored for clarity
(otherwise, since the number of waiting steps $l$ 
is finite in the present 2D architecture, 
we can take the memory errors into account 
by replacing the error probability 
$p_g$ of the two-qubit gates
with $p_g + \eta l $,
where $\eta$ indicates
the memory error probability per step).  
Such QALUs are connected via quantum channels of fidelity $F$.
Here the channel fidelity $F$ means that we can share
a maximally entangled state (MES) of fidelity $F$ between
the nearest-neighbour QALUs.
The remote entangling operations can be used for this purpose \cite{RemoteEntangle}.
The quantum channels are assumed to be relatively noisy ($F\sim 0.7$--$0.9$),
which relaxes the complexity in scaling up the size of the architecture. 

\section{Entanglement pumping and teleportation-based two-qubit gate}
The noise introduced by the quantum channel
degrades the accuracy of the gate operation
between the separate systems.
In classical computation,
this kind of obstacles can be overcome by using the multiplexing technique \cite{vonNeumann52,Han03},
where multiple copies of data with majority voting 
are used to ensure the reliability of logic circuits.
In the case of quantum physics, however,
we cannot make copies of an unknown quantum state
due to the no-cloning theorem  \cite{Wooters82}.
Instead, we can use entanglement purification \cite{Purification1,Purification2} 
and quantum teleportation \cite{GC99} 
in order to improve the fidelity of the gate operations between two QALUs.
In particular, 
a novel method is devised
to realise gate operations of high accuracy,
where an intelligent use of the four-qubit local system
allows us to obtain MES of high fidelity
even with very noisy quantum channels,
as shown below.

\begin{figure}
\centering
\includegraphics[width=80mm]{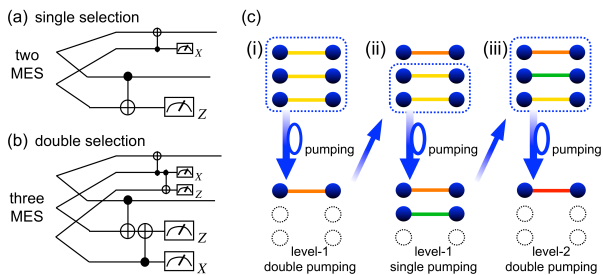}
\caption{
Entanglement purification with entanglement pumping. (a) The entanglement purification with 
single selection \cite{Purification1,Purification2}. (b) The entanglement purification with double selection \cite{FY09}.
(c) (i) The level-1 entanglement pumping with double selection,
where one noisy MES is purified by using two noisy MESs as ancillae.
(ii) The level-1 entanglement pumping with single selection,
where one noisy MES is purified by using one noisy MES. 
(iii) The level-2 entanglement pumping with double selection,
where the output state of (i) is purified 
by using the output state of (ii) and one noisy MES as ancilla.
}
\label{purification}
\end{figure}
The noisy copies of the MES (we call it ``noisy MES" hereafter), 
which are shared by using the noisy quantum channels,
are purified by using the entanglement pumping scheme \cite{RepeaterQIP,Pumping} as follows [see Fig. \ref{purification} (c)]:
(i) A noisy MES
(we call it the target pair)
is purified by reducing its bit-flip ($X$) error via
level-1 entanglement pumping with double selection \cite{FY09}, 
where two noisy MESs are used as the ancillae.
The phase-flip ($Z$) error in the target pair increases in this process.
(ii) Another noisy MES is purified by using level-1 entanglement pumping with single selection \cite{Purification1,Purification2}, where one noisy MES is used as the ancilla.
Here the target pair is untouched.
(iii) The target pair is purified by reducing its $Z$ error via
level-2 entanglement pumping with double selection, where the successful output state of (ii) and one noisy MES are used as the ancillae.
Here the increase in the $X$ error is kept small thanks to the level-1
pumping in step (ii).
At each step, we repeat the pumping process 
from a few to several times.
Then, we finally obtain the level-2 double-pumped MES of high fidelity.
\begin{figure*}
\centering
\includegraphics[width=130mm]{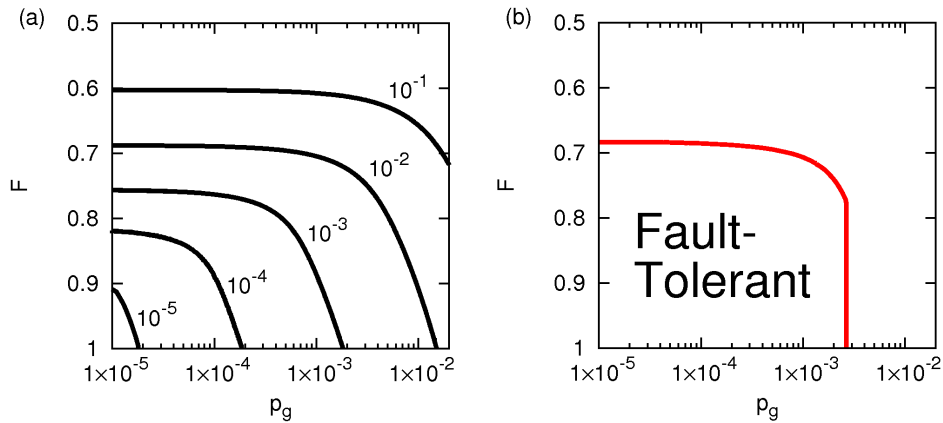}
\caption{
Infidelities of the purified MES and threshold curve.
(a) The contour plot of the output infidelities $1- \bar F$ with respect to
the channel fidelity $F$
and the error probability $p_g$ of local operations,
where $p_{ij} = p_g/15$ and $p_M =p_g$ are specifically adopted. 
(b) The threshold curve with respect to the channel fidelity $F$
and the error probability of the local operations $p_g=p_M$.
}
\label{Fidelity}
\end{figure*}	

The noisy MES $\rho _{\rm in}$,
which is shared by the noisy quantum channel,
is described as
\begin{eqnarray}
\rho _{\rm in} = \sum _{i=0}^{3} F_{i} (\sigma _{i} \otimes \sigma _0) |\phi \rangle \langle \phi | (\sigma _{i} \otimes \sigma _0),
\end{eqnarray}
where $|\phi\rangle= (|00\rangle+|01\rangle+|10\rangle-|11\rangle)/2$. 
The purification processes with single and double selections 
in the above pumping scheme can be described as
maps
$ F'_{l}=S_l^{ij} F_i ^{(1)} F_j^{(2)}/p_s$
and
$ F'_{l}=D_l^{ijk} F_i ^{(1)} F_j^{(2)} F_k^{(3)}/p_d$, respectively.
Here, $F'_l$ and $F^{(n)}_i$ are the output and
the $n$th input fidelities, respectively,
and $p_s$ and $p_d$ are the normalization factors (i.e. $\sum _{l}F'_l =1$)
meaning the success probabilities of the purification processes.
Starting from the channel fidelity $(F \equiv F_0, F_1,F_2,F_3)$,
the fidelity of the level-2 double-pumped MES 
$(\bar F \equiv \bar F_0, \bar F_1, \bar F_2 , \bar F_3 )$
is calculated by using the tensors $S_l^{ij}$ and $D_l^{ijk}$,
which can be written in terms of $p_{ij}$ and $p_M$ 
(see Appendix A).
The contour plots of the output infidelities $1-\bar F$ with respect to
the channel fidelity $F$
and the error probability $p_g$ of the local operations
are shown in Fig. \ref{Fidelity} (a),
where $F_{1,2,3}=(1-F)/3$, $p_{ij} = p_g/15$ and $p_M =p_g$ are  adopted specifically.
Compared to the single pumping \cite{Pumping,Jian07},
the output fidelity is significantly improved
in the present scheme with double selection, 
without increasing the number of spatial resources.

By using the purified MES of high fidelity,
we perform the teleportation-based two-qubit gate (TTG) between
the data qubits stored in the nearest-neighbour QALUs.
Including the memory space for the data qubit, 
a total of four qubits (i.e. three auxiliary qubits for the entanglement pumping and one data qubit) 
are required in each QALU.
(The four-qubit system seems to be the minimum for our purpose,
since we cannot purify both bit- and phase-flip noise by using only two auxiliary qubits.)
The probabilities $\bar p_{ij}$ of the $\sigma _i \otimes \sigma _j$ errors
after the TTG can be calculated
in terms of $\bar F _{i}$, $p_g$ ($p_{ij}=p_g/15$), and $p_M$.
Roughly speaking, the gate infidelity is given by 
$ \sum _{(i,j) \neq (0,0)} \bar{p}_{ij} \simeq  (1 - \bar F)+  2 p_g + 2 p_M $,
which can be understood from the fact that
the TTG is implemented with
one purified MES,
two local two-qubit gates
and measurements \cite{GC99} (see Appendix B for details).

\section{Topological fault-tolerant quantum computation}
As shown above, we can perform two-qubit gates of high accuracy
between the nearest-neighbour two QALUs in the teleportation-based way,
but they are still subject to small error.
In order to handle this
and achieve gate operations of arbitrary accuracy,
we perform fault-tolerant TQC \cite{Raussendorf07a},
where the surface code protects quantum information
by virtue of the topological degeneracy \cite{Kitaev}. 
The logical CNOT gate operations in TQC
are implemented by braiding anyons (defects) on the surface. 
Universal fault-tolerant quantum computation can be achieved by using the state injection
and the magic state distillation \cite{Raussendorf07a,Magic}.
Actually, these operations for fault-tolerant universal quantum computation 
can be executed by using only 2D nearest-neighbour two-qubit gates
and single-qubit measurements \cite{Raussendorf07a}, and therefore
TQC is best suited for the present distributed architecture.

\begin{figure*}
\centering
\includegraphics[width=120mm]{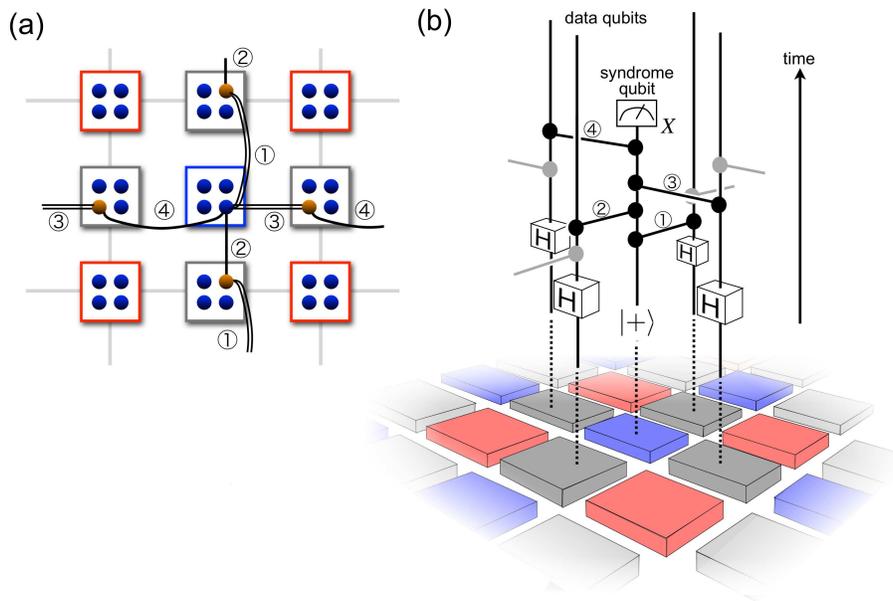}
\caption{
Syndrome measurements for topological error correction. 
(a) The syndrome measurement 
for topological error correction. The TTGs are impelented in numerical order ($1 \rightarrow 2 \rightarrow 3 \rightarrow 4$).  The single line which connects the QALUs 
indicates the teleportation-based CZ gate. The double line indicates the teleportation-based CZ gate preceded by the Hadamard operation, where the Hadamard operation is incorporated in the TTG (see suplemental material). 
By measuring the qubit on the blue QALU,
one can obtain the eigenvalue of the stabilizer operator $X^{\otimes 4}$. 
(b) The circuit diagram in 3D (space-like 2D and time-like 1D) of the syndrome measurement.
}
\label{topological}
\end{figure*}

The noise threshold of TQC is determined 
from the error correction procedures,
which are implemented repeatedly in the bulk region
of the surface code \cite{Raussendorf07a}. 
The syndrome measurement for
the topological error correction can be executed as shown in Fig. \ref{topological} (a) and (b),
where the QALUs of gray and blue (red) squares indicate
the data and $X$ ($Z$) syndrome QALUs, respectively.
After performing the TTGs,
the blue qubits are measured in the $X$ basis,
whose outcome corresponds to the eigenvalue
of the stabilizer operator $X^{\otimes 4}$
and is used to correct $Z$ errors [see Fig. \ref{topological} (b)].
The $X$ errors are transformed to $Z$ errors
by the Hadamard operations, which is incorporated into the TTGs,
and corrected in the same way at the next 
step by using the $Z$ syndrome qubits (red boxes).

The errors during the error correction
can be characterised by the independent and correlated 
errors on the three-dimensional (i.e. space-like 2D and time-like 1D)
lattice \cite{Raussendorf07a}.
In Ref. \cite{Raussendorf07a},
the threshold values of the 
independent and correlated error probabilities
$q_{\rm ind}$ and $q_{\rm cor}$, respectively,
have been numerically estimated as 
$(q_{\rm ind},q_{\rm cor})=(2.3\%, 0.40\%)$,
which are attributed to 
the preparation, gate, measurement errors
with equal probability $0.75 \%$.
In our case,
these probabilities
$q_{\rm ind}$ and $q_{\rm cor}$ are calculated
in terms of $\bar F_{2,3}$, $p_g$ and $p_M$ as 
(see Appendix C for the details)
\begin{eqnarray}
q_{\rm ind}&=&
 4(\bar F_2+\bar F_3)+\frac{40}{15}p_g +p_M ,
\\
q_{\rm cor} &=& \frac{8}{15}p_g+p_M.
\end{eqnarray}
Thus if they satisfy $q_{\rm ind}<2.3 \%$ and $ q_{\rm cor}<0.40 \%$,
fault-tolerance of the present architecture is ensured
(the true threshold will be slightly higher than 
the value calculated from the above condition,
which can be determined by a full numerical simulation).
By using these conditions and the output fidelity $\bar F_{1,2,3}$
obtained in the previous section,
we calculate the fault-tolerant region of the channel fidelity $F$
and the error probability $p_g$ of local operations,
where $p_M=p_g$ is taken for simplicity.
The threshold curve for these physical parameters $(p_g,F)$
is plotted in Fig. \ref{Fidelity} (b).
Specifically, with $F \sim 1$,
the threshold value of local operations 
is obtained as $p_g=0.26\%$ ($p_g=0.5\%$ when $p_M=4p_g/15$ is adopted \cite{Knill05,FY10}).
It is also seen that
the present architecture works even 
with very noisy quantum channels of fidelity $F \sim 0.7$
provided the error probability of the local operations are reasonably small $p_g \sim 0.1\%$.
This result contrasts with the requirements, $F \sim 0.95$ and $p_g\simeq 10^{-4}$, 
in the 1D architecture with five-qubit distributed systems \cite{Jian07}.

The computational overhead
required in the present architecture can be quantified
by the total amount $R$ of local operations
plus quantum communication (i.e. the number of noisy MESs).
The overhead $R$ is given by $R = K T$,
where $T$ indicates the number of two-qubit gates in TQC, and
$K$ is the total amount of local operations plus
quantum communication (i.e. the number of initial MESs) per TTG.
Specifically,
the overhead $K$ is a few tens (a few hundreds) 
when $F\sim 0.9$ (0.7) and $p_g \sim 0.1\%$,
which mainly depends on the channel fidelity $F$.
In order to factorise an $n$-bit composite number,
we need $40n^3$ Toffoli gates \cite{SemiconductorQC},
each of which is implemented by using seven $\pi/8$ gates.
Thus a total of $300n^3$ $\pi/8$ gates are required 
to factorise a $n$-bit composite number.
This means that each $\pi /8$ gate has to work 
with an error probability of $\sim 1/(300n^3)$.
When $n=1024$, 
such a $\pi/8$ gate requires $2\times 10^{10}$ physical two-qubit gates 
under the physical error probability at 1/3 of the topological threshold,
and therefore a total of 
$ T \sim 6 \times 10^{21}$ two-qubit gates are required in TQC.
On the other hand,
$K \sim 40$ is required to achieve 1/3 of the topological threshold 
with $F \simeq 0.9$ and $p_g \simeq 0.1\%$ (see Appendix D).
As a result, the total overhead for the present architecture 
amounts to be $R=KT\sim 2 \times 10 ^{ 23}$.

Although the above quantum overhead is large, comparable 
complexity is achieved in today's classical computer 
within $10^5$ sec (a few tens of hours),
where $7 \times 10^8$ transistors integrated in a central processing unit (CPU)
work at rates of $3\times 10^9$Hz \cite{Intel} (i.e. $2 \times 10^{23} \simeq 7 \times 10^8 \times  3 \times 10^9 \times 10^5$).
It contrasts with the fact that the recent factorization of a general
768-bit composite number has taken
an overhead equivalent to 1677 years with a CPU of 2.2GHz \cite{768bit}.

\section{Physical implementation}
All key technologies required in the present architecture
have already been demonstrated experimentally 
in various physical systems, such as trapped ions \cite{iontraps}
and NV-centres in diamond \cite{NVcenter}.
Among them, state-of-the-art technologies in
the trapped ion systems have already 
achieved highly accurate controls.
Multiple two-qubit gates, single qubit rotations, and readouts
have been achieved with error probabilities 
$7\times 10^{-3}$ \cite{IonTwoGate}, $2 \times 10^{-5}$ \cite{IonSingleGate}, and $10^{-4}$ \cite{IonReadout,IonReadout2}, respectively.
These benchmarks are comparable to the requirements
for the local operations in the QALUs.
In order to attain scalability, the ion qubits could be distributed
over separate trap zones.
There are two main schemes, 
which can be used as the quantum channels,
to entangle separately trapped ion qubits.
One is based on the quantum charge-coupled device \cite{QCCD}
with the microfabricated ion trap technologies,
where entangling operations are implemented
by shuttling ion qubits between the storage and interaction regions.
The transport of ion qubits has succeeded experimentally
with high accuracy,
and a two-qubit gate has been demonstrated
with high fidelity $\sim 0.9$ \cite{IonTransport2,IonTransport3}.
Another approach utilises photons as flying qubits in order to entangle 
two separate ion qubits,
where the fidelity $\sim 0.9$ has been achieved experimentally 
\cite{IonEntangling}.
In order to couple ion qubits efficiently with photons,
surface-electrode ion traps integrated with
microscale optics have also been investigated recently \cite{IonMicroOpt1,IonMicroOpt2}.
These fidelities of the above two entangling operations 
are well above our requirement.
($p_g = 5 \times 10^{-3}$, $p_M=10^{-4}$ and $F=0.9$ leads
to the values of $q_{\rm ind}$ and $q_{\rm cor}$ below the thresholds.)

Solid state systems are another promising
candidates for a physical implementation of the present architecture. 
In the NV-centre diamond systems, particularly,
there are well-defined local systems as the QALUs \cite{NVsystem,NVcenter}.
The control of the 12--16 dimensional systems (electric spin-1 and two or three nuclear spin-1/2) \cite{NVcenter,NVcenterb} and the single-shot readout \cite{NVreadout,NVreadoutb}
have already been achieved.
The entanglement between 
the polarization of a single optical photon and
a single electronic spin in a NV-centre diamond \cite{NVentangle},
and two-photon interference from separated NV-centers \cite{NVentangleb,NVentanglec}
have also been observed recently.
These technology can be used as the quantum channel for the present 
distributed architecture.
Furthermore,
universal dynamical decoupling \cite{NVdecoupling} could be 
employed to suppress noise
and achieve reasonably accurate local operations
required for the present architecture.

\section{Discussion and conclusion}
We have proposed a distributed architecture
for scalable quantum computation.
The present architecture
works with reasonably accurate small (four-qubit) local systems,
which are connected by very noisy quantum channels.
All key ingredients employed in the present architecture
have  already been experimentally demonstrated,
and the accuracies required for them are comparable
to the recent experimental achievements.
Actually this is the first proposal of
a practically scalable architecture which works well
even with the preexisting level of the experimental devices.
We believe that this proposal
fills the gap between top-down and bottom-up
approaches towards practically scalable quantum computation
and gives a good guideline and benchmark in the development
of quantum devices.

Finally let us mention that
there are a lot of rooms to improve
the performance of the present architecture.
Instead of the bipartite entangled states,
we can also use multipartite entangled states
as resource states for teleportation-based multi-qubit gates.
Since a part of errors during local operations,
which are to be performed in the future computation,
are removed beforehand through purification,
the physical threshold would be improved.
Furthermore, by using improved decoding algorithm,
we can fully utilise the potential power of the surface code.
It boosts the topological threshold from $0.75\%$ to $1.1$--$1.4\%$ \cite{Wang11}.
The topological colour codes \cite{Bombin},
whose threshold value $4.8\%$ \cite{Andrist11}
is higher than that ($3.2\%$) of the Kitaev's surface code \cite{Kitaev},
can be also implemented on the present architecture.
These upgrades will improve the threshold values
of the local operations to $\sim 1\%$, which will become comparable to
that of the non-distributed fault-tolerant schemes
without modular scalability \cite{Knill05,Raussendorf07a,FY10}. 

\section*{Acknowledgements}
This work was supported by the Funding Program for World-Leading Innovative R \& D on Science and Technology (FIRST), MEXT Grant-in-Aid for Scientific Research on Innovative Areas 20104003 and 21102008, the MEXT Global COE Program and MEXT Grant-in-Aid for Young scientists (A) 23684035.

\appendix
\section{Entanglement pumping}
\subsection{Entanglement pumping with single selection}
In the entanglement purification with single selection,
one noisy MES is purified by using one noisy MES
as the ancilla as shown in Fig. 2 (a).
Assuming that all errors are given 
by probabilistic Pauli errors, 
the purification map can be described as 
\begin{eqnarray}
F_{k}' = S_{k}^{ij}F_{i}^{(1)}F_{j}^{(2)},
\end{eqnarray}
where $F'_k$ indicate the fidelities of the output state.
The transition probability tensor $S^{ij}_{k}$ is given by
\begin{equation}
S^{ij}_{k} =\sum _{l=0,3} {M}^{a}_{l} G^{ij}_{ka} \;,
\end{equation}
where $M^a_l$ and $G^{ij}_{ka}$ correspond to the bilateral measurement error 
and bilateral CNOT gate followed by the two-qubit gate errors \cite{FY09},
respectively,
and the summation $\sum _{l=0,3}$ means the postselection
according to the bilateral measurement outcomes
as shown in Fig. \ref{Purification} (a),
which are designed to check bit-flip errors.
Throughout the level-1 single pumping, 
the initial noisy MES is used as the ancilla for pumping, and hence
we take  $\mbF_{j}^{(2)}=\mbF^{\rm ini} \equiv (F,\frac{1-F}{3},\frac{1-F}{3},\frac{1-F}{3})$ \cite{Jian07}.
Then, the purification can be viewed as a map from $\mathbb{R}^4 \rightarrow 
\mathbb{R}^4$,
$\mbF'=\mathcal{S}_1(\mbF ^{(1)})$.
Starting with the initial noisy MES (i.e. $\mbF^{(1)}=\mbF^{\rm ini}$), 
the single pumping $\mathcal{S}_1$ is repeatedly applied $n_1$ times.
As a result, we obtain the level-1 single-pumped MES of the fidelity 
\begin{eqnarray}
\mbF ^{\rm Lv1}  = \mathcal{S}_1 ^{n_1} (\mbF^{\rm ini})/p_{\rm Lv1} \;,
\end{eqnarray}
where $p_{\rm Lv1} \equiv \sum _{k=0}^{3}  [\mathcal{S}_1 ^{n_1} (\mbF^{\rm ini})]_k$
indicates the net success probability of the level-1 single pumping.
\setcounter{figure}{4}
\begin{figure}
\centering
\includegraphics[width=80mm]{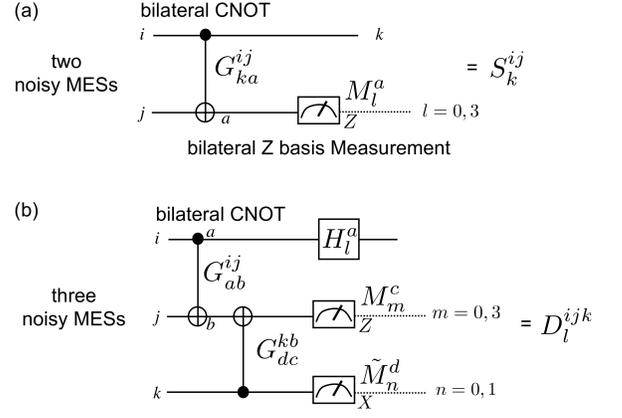}
\caption{(a) Entanglement purification with single selection. (b) Entanglement purification with double selection.}
\label{Purification}
\end{figure}

\begin{figure}
\centering
 \includegraphics[width=70mm]{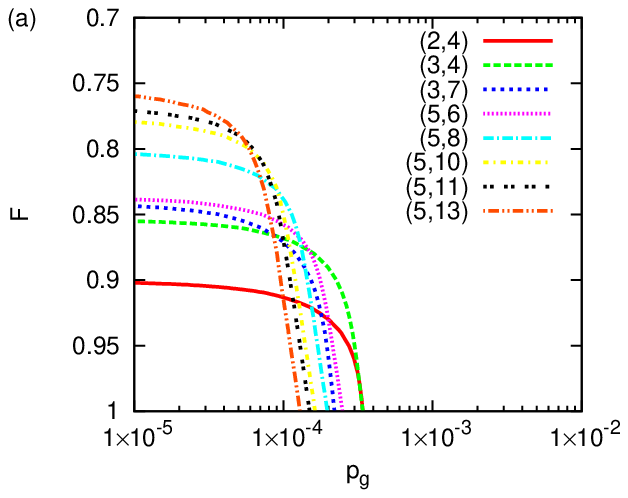}
 \\
 \includegraphics[width=70mm]{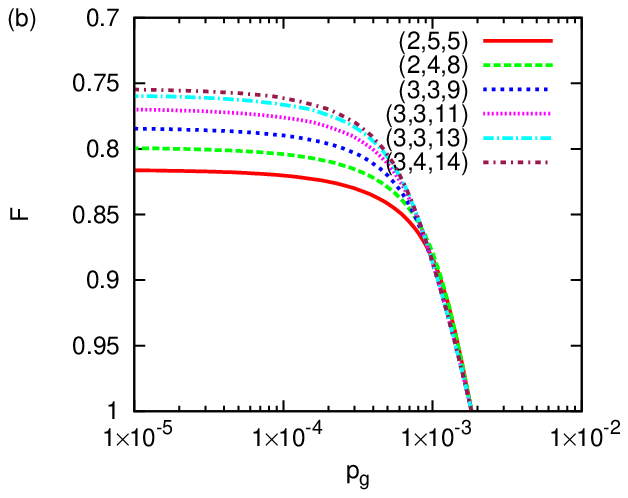}
\caption{Infidelities of the purified MESs. 
(a) The contour of the infidelity $1-\bar F = 10^{-3}$ of the level-2 single-pumped MES is plotted with respect to the channel fidelity $F$ 
and error probability $p_g=p_M$ of local operations for each $(n_1,n_2)=(2,4),(3,4),(3,7),(5,6),(5,8),(5,10),(5,11), (5,13)$. (b) The contour of the infidelity $1-\bar F = 10^{-3}$ of the level-2 double-pumped MES is plotted with respect to the channel fidelity 
$F$ and error probability $p_g=p_M$ of the local operations 
for each $(n_1,m_1,m_2)=(2,5,5),(2,4,8),(3,3,9),(3,3,11),(3,3,13),(3,4,14)$.
}
\label{PumpFidelity}
\end{figure}

In the level-2 single pumping  \cite{Jian07}, 
the output state of the level-1 single pumping
is further purified by using the level-1 single-pumped MES as the ancilla, 
where the phase-flip errors are checked.
The purification map $\mathcal S_2$ can be described as
\begin{eqnarray}
F_{k}' = \tilde S_{k}^{ij}F_{i}^{(1)}F_{j}^{\rm Lv1}\; ,
\end{eqnarray}
where $\tilde S^{ij}_{k}  = H^{c}_{k} S^{ab}_{c}H^{i}_a H^{j}_{b}$,
with $H^{i}_{j}$ being the bilateral Hadamard transformation.
By repeatedly applying the level-2 single pumping $n_2$ times, 
we obtain the level-2 single-pumped MES of the fidelity
\begin{eqnarray}
\mbF ^{\rm Lv2}  = \mathcal{S}_2 ^{n_2} (\mbF^{\rm Lv1} )/p_{\rm Lv2} \;.
\end{eqnarray}
The contour of the infidelity $1- \bar F = 10^{-3}$ 
of the level-2 single-pumped MES is plotted with respect to the
channel fidelity $F$ and error probability $p_g=p_M$ of local operations 
for each $(n_{1},n_{2})=(2,4),(3,4),(3,7),(5,6),(5,8),(5,10),(5,11), (5,13)$ in Fig. \ref{PumpFidelity} (a).
There is a tradeoff between the channel fidelity $F$ and error probability $p_g=p_M$ of local operations when the the numbers of repetitions $(n_1,n_2)$ are increased.
This behavior can be understood that 
either bit or phase error is checked repeatedly at each pumping level,
which is designed to obtain the output MES of high fidelity 
even with low channel fidelity $F$ as discussed in Ref. \cite{Jian07}.

\subsection{Entanglement pumping with double selection}
In the level-1 entanglement pumping with double selection,
one noisy MES is purified by using two initial noisy MESs as the ancillae as shown in Fig. \ref{Purification} (b).
Similarly to the previous case,
the purification map $\mathcal{D}_{1}$
can be described as
\begin{eqnarray}
F' _{l}= \sum_{m=0,3 ; n=0,1} {D}^{ijk}_{lmn}F_i^{(1)} F_j ^{\rm ini} F_k ^{\rm ini} ,
\end{eqnarray}
where the transition probability tensor is given by ${D}^{ijk}_{lmn}
= H^{a}_{l} M^{c}_{m} \tilde{M}^{d}_{n} G^{kb}_{dc} G^{ij}_{ab} $
with $\tilde{M}^{d}_{n}=H^{j}_{n} M^{i}_{j} H^{d}_{i}$.
Starting with the initial noisy MES (i.e. $\mbF^{(1)}=\mbF^{\rm ini}$),
the level-1 double pumping $\mathcal{D}_1$ is repeatedly applied $m_1$ times, 
and then we obtain the level-1 double-pumped
MES of the fidelity
\begin{eqnarray}
\tilde \mbF^{\rm Lv1} = \mathcal{D}_1 ^{m_1} (\mbF^{\rm ini})/ r _{\rm Lv1} \;,
\end{eqnarray}
where $ r _{\rm Lv1}$ is the net success probability similarly to the previous case
with single selection.
In the level-2 double pumping,
the output state of the level-1 double pumping
is further purified by using the level-1 single pumped MES and initial noisy MES as the ancillae.
The level-2 double pumping $\mathcal{D}_2$ can be described as
\begin{eqnarray}
\mbF' = \sum_{m=0,3 ; n=0,1} {D}^{jkl}_{imn}F_j^{(1)} F_k ^{\rm  Lv1} F_l ^{\rm ini}.
\end{eqnarray}  
By applying $\mathcal{D}_2$ repeatedly $m_2$ times,
we obtain the level-2 double-pumped MES of the fidelity
\begin{eqnarray}
\tilde \mbF^{\rm Lv2}= \mathcal{D}_{2} ^{m_2} ( \tilde \mbF ^{\rm Lv1})/r_{\rm Lv2} \;.
\end{eqnarray}
The contour of the infidelity $1-\bar F=10^{-3}$
is plotted with respect to the channel fidelity $F$ and
error probability $p_g=p_M$ of the local operations for each 
$(n_1,m_1,m_2)=(2,5,5),(2,4,8),(3,3,9),(3,3,11),(3,3,13),(3,4,14)$
in Fig. \ref{PumpFidelity} (b).
In Fig. 3 (a), the result with $(n_1,m_1,m_2)=(3,4,14)$ is plotted.
In the level-2 double pumping,
the acceptable range of the channel fidelity $F$ is improved without sacrificing that of
the error probability $p_g$ of local operations,
when the numbers of repetitions $(n_1,m_1,m_2)$ are increased.
This is due to the following two factors:
(i) Both bit and phase errors are checked at each pumping level. 
(ii) A large amount of operational errors,
which are left on the output state of single selection,
can be detected by double selection \cite{FY09}.

\section{Error analysis of TTGs}
We employ three types of TTGs in the syndrome measurement
(see Fig. 4 (a)).
\subsection{TTG of type I}
The TTG of type I [\textcircled{\footnotesize 1} in Fig. 4 (a)] is as follows:
\begin{eqnarray}
\centering
\includegraphics[width=80mm]{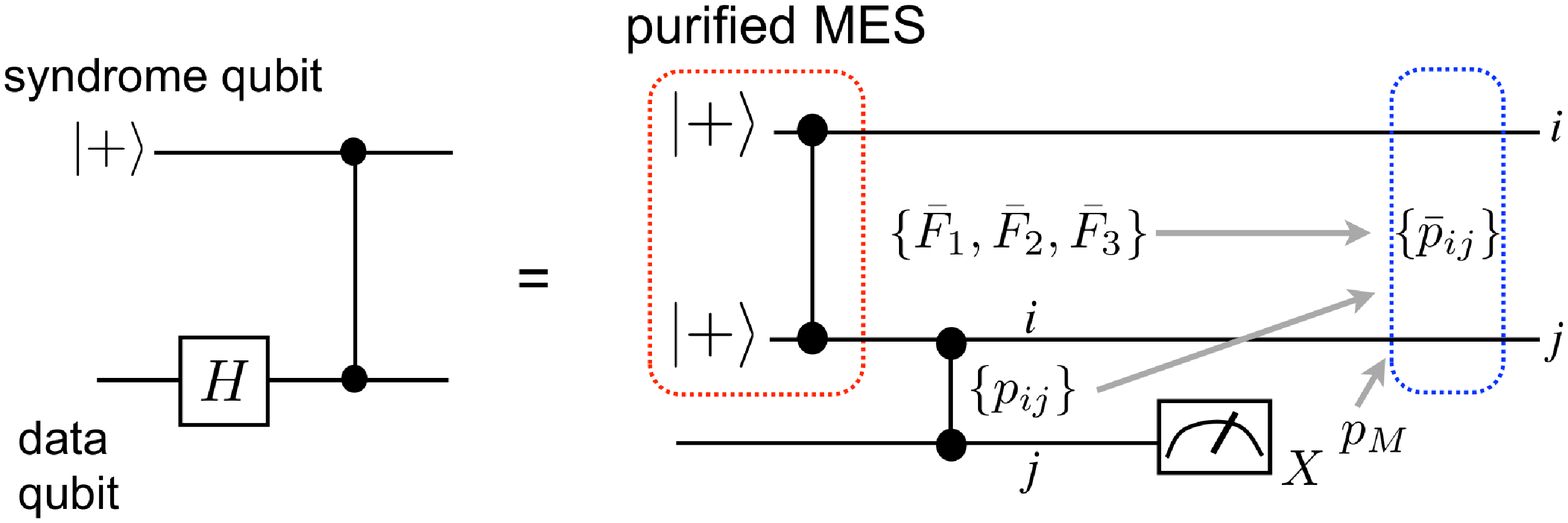}
\end{eqnarray}
where the Hadamard operations is required to transform $Z$ errors
to $X$ errors.
The probabilities of $\sigma _i \otimes \sigma _j$ errors on the two output qubits
are given by 
{\small
\begin{eqnarray*}
\begin{array}{c||c|c|c|c}
i  \backslash j  & I & X & Y & Z
 \\
 \hline \hline
I &&p_{X \{I,X\}}& p_{Y \{I,X\}}&  \bar F _{1}+p_{Z \{I,X\}} 
 \\
 \hline
X & & & &
 \\
 \hline
Y & & & &
 \\
 \hline
Z & \bar F_{3} +p_{X \{Y,Z\}}& p_M+p_{I \{Y,Z\}} &p_{Z \{Y,Z\}} & \bar F_{2} +p_{Y \{Y,Z\}}
\end{array}
\end{eqnarray*}
}
where we switched the notation from $p_{ij}$ to $p_{AB}$ ($A,B \in \{ I,X,Y,Z\}$), and $p_{A \{B,C\}}$ means $p_{AB} + p_{AC}$.

\subsection{TTG of type II}
The TTG of type II [\textcircled{\footnotesize 2} and \textcircled{\footnotesize 4} in Fig. 4 (a)] is as follows:
\begin{eqnarray}
\centering
\includegraphics[width=80mm]{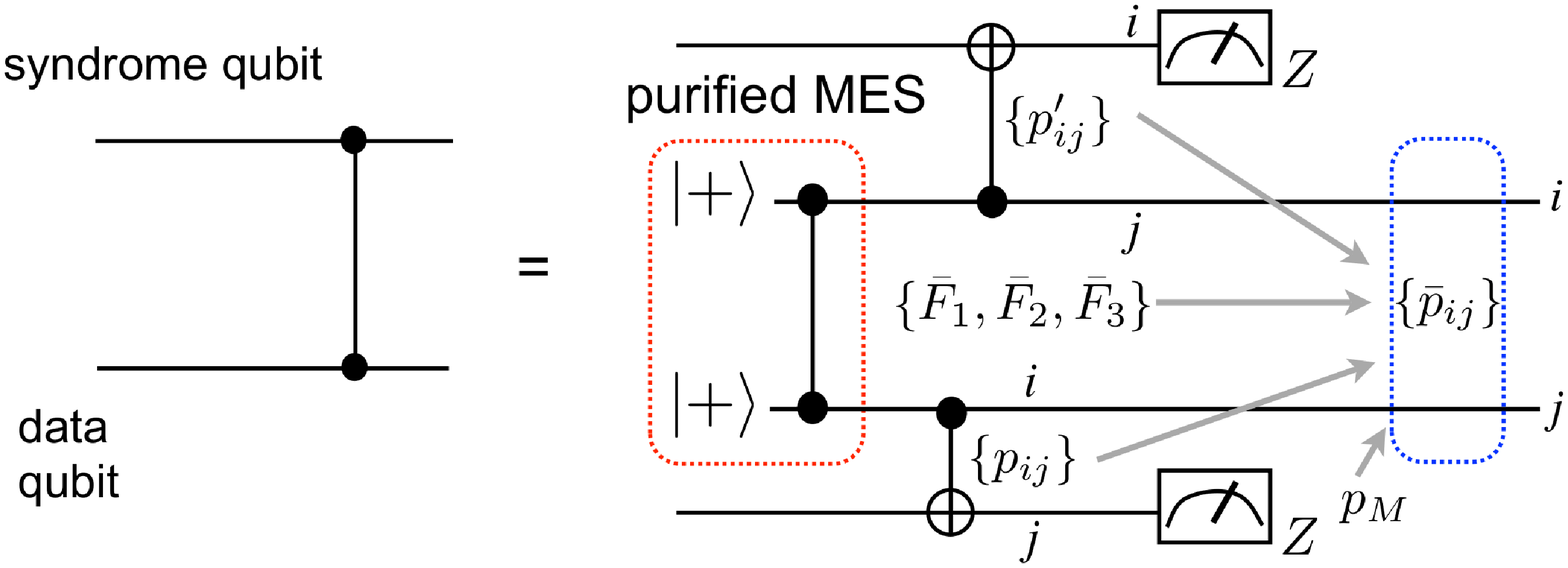}
\end{eqnarray}
The probabilities of $\sigma _i \otimes \sigma _j$ errors on the two output qubits
are given by 
{\small
\begin{eqnarray*}
\begin{array}{c||c|c|c|c}
i  \backslash j  & I & X & Y & Z
 \\
 \hline \hline
I &   &p_{X \{I,Z\}}& p_{Y \{I,Z\}}&  \bar F _{1}+p_{Z \{I,Z\}} 
 \\
  &   & &  & +p' _{\{X,Y\}X}
 \\
 \hline
X &p'_{\{I,Z\}X} & & & p_M+ p' _{\{X,Y\}I}
 \\
 \hline
Y & p'_{\{I,Z\}Y}& & & p' _{\{X,Y\}Z}
 \\
 \hline
Z & \bar F_{3} +p_{X \{X,Y\}}& p_M+p_{I \{X,Y\}} &p_{Z \{X,Y\}} & \bar F_{2} +p_{Y \{X,Y\}}
\\
& + p'_{\{I,Z\}Z} & & &+p' _{\{X,Y\}Y}
\end{array}
\end{eqnarray*}
}
where $p_{ij}$ and $p'_{ij}$ indicate the error probabilities of the top and bottom two-qubit gates, respectively, as shown in the above diagram.

\subsection{TTG of type III}
The TTG of type III [\textcircled{\footnotesize 3} in Fig. 4 (a)] is as follows:
\begin{eqnarray}
\centering
\includegraphics[width=80mm]{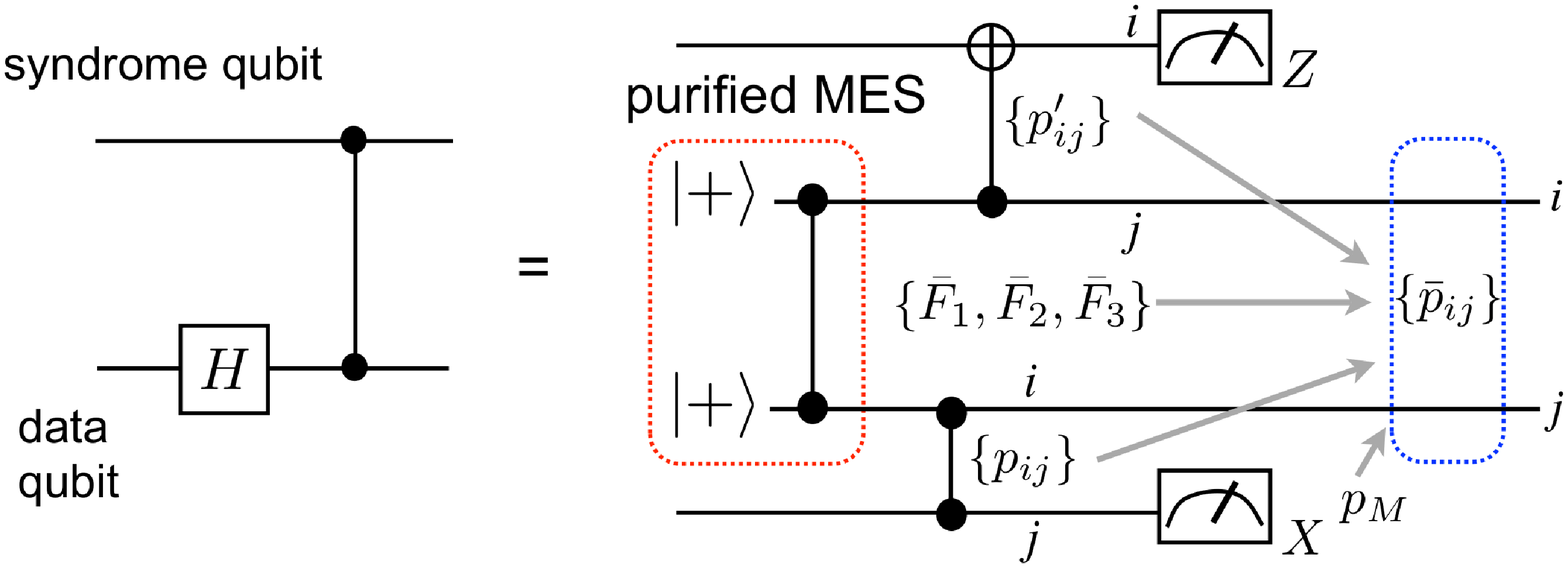}
\end{eqnarray}
The probabilities of $\sigma _i \otimes \sigma _j$ errors on the two output qubits
are given by 
{\small
\begin{eqnarray*}
\begin{array}{c||c|c|c|c}
i  \backslash j  & I & X & Y & Z
 \\
 \hline \hline
I &    &p_{X \{I,X\}}& p_{Y \{I,X\}}&  \bar F _{1}+p_{Z \{I,X\}} 
 \\
 &    && &+p' _{\{X,Y\}X}
\\ 
 \hline
X &p'_{\{I,Z\}X} & & & p_M+ p' _{\{X,Y\}I}
 \\
 \hline
Y & p'_{\{I,Z\}Y}& & & p' _{\{X,Y\}Z}
 \\
 \hline
Z & \bar F_{3} +p_{X \{Y,Z\}}& p_M+p_{I \{Y,Z\}} &p_{Z \{Y,Z\}} & \bar F_{2} +p_{Y \{Y,Z\}}
\\
 &+ p'_{\{I,Z\}Z}& & & +p' _{\{X,Y\}Y}

\end{array}
\end{eqnarray*}
}
where $p_{ij}$ and $p'_{ij}$ indicate the error probabilities of the top (CNOT) and bottom (CZ) two-qubit gates, respectively,
as shown in the above diagram.

\section{Threshold analysis}
\begin{figure}
\centering
\includegraphics[width=80mm]{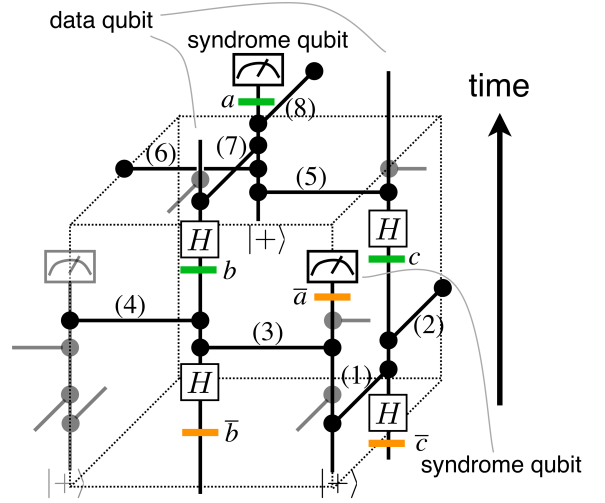}
\caption{The unit cell of the syndrome measurements for topological error correction.
The two-qubit gates are implemented in the numerical order.
The errors during the error correction on the primal and dual lattices 
can be described as $Z$ errors 
located at $a,b,c$ (green lines) and $\bar a, \bar b, \bar c$ (orange lines), respectively.
}
\label{RaussendorfLattice}
\end{figure}

The syndrome measurements for topological error correction are implemented as shown in Fig. \ref{RaussendorfLattice}.
The errors during the error correction on the primal lattice \cite{Raussendorf06,Raussendorf07a,Raussendorf07b} can be described as $Z$ errors located at $a$, $b$, and $c$, which are depicted with green lines in Fig. \ref{RaussendorfLattice}.
The errors at $a$ and $b,c$ correspond to errors on the measured syndrome
and data qubits, respectively.
Similarly, the errors during the error correction on the dual lattice
are located at $\bar a$, $\bar b$, and $\bar c$,
which are depicted with orange lines in Fig. \ref{RaussendorfLattice}.
Since the error corrections on the primal and dual lattices are implemented independently,
we can ignore the correlations between errors on the primal and dual lattices
(i.e. correlated errors between $a,b,c$ and $\bar a , \bar b , \bar c$).
Furthermore,
behavior of the errors on the dual lattice is equivalent to
that on the primal lattice,
since they are symmetric under reflection and rotation.
Thus we consider only the errors on the primal lattice, 
i.e., those errors at $a$, $b$, and $c$.

Let $q^{a}$, $q^{b}$, and $q^{c}$ denote
the probabilities of the independent $Z$ errors at $a$, $b$, and $c$,
respectively.
If errors are located on the data qubits independently with equal probability,
and syndrome measurements are perfect, 
then $q^a=0$, $q^b=q^c$.
In such a case, the threshold value is given by $q^b=q^c=11\%$ \cite{Kitaev}.
With $q^a=q^b=q^c$ (i.e. independent errors on the syndrome and data qubits
with equal probability),
the threshold values have been estimated as $q^a=q^b=q^c=2.9\%$ 
\cite{Wang} and $3.3\%$ \cite{Ohno}
by using the minimum-weight perfect matching algorithm
and the random-plaquette $Z_2$ gauge theory, respectively.

In the present case (and also in Ref. \cite{Raussendorf07a,Raussendorf07b}), 
however, we have to take the correlated errors into account,
since the two-qubit gates for the syndrome measurements 
introduce correlation.
By using the commutation relations between errors and two-qubit gates,
it is found that the correlated errors are located only at $(a,b)$ and $(a,c)$ of the same unit cell 
and $(b,b)$ of the neighboring unit cells.
Such probabilities are denoted by $q^{a,b}$, $q^{a,c}$, and $q^{b,b}$.
The probabilities of independent and correlated errors 
can be calculated in terms of the error probabilities $p_{AB}^{(l)}$
of the $l$th two-qubit gate (see Fig. \ref{RaussendorfLattice}),
where $A$ and $B$ indicate errors on the syndrome and data qubits respectively,
and the Hadamard operation is included in the two-qubit gate for $l=1,3,5,7$:
\begin{eqnarray}
q^{a} &=& p^{(5)}_{z\bar x}+p^{(6)}_{z\bar x}+p^{(7)}_{z\bar x}+p^{(8)}_{z\bar x}+p_P+p_M,
\label{eq1}\\
q^{b} &=&p^{(3)}_{\bar x z} + p^{(3)}_{x \bar z } 
+p^{(4)}_{x z} + p^{(4)}_{\bar x z}
+p^{(7)}_{zx} +p^{(8)}_{\bar z x},
\\
q^{c} &=&p^{(1)} _{\bar x z} +p^{(1)}_{x \bar z }+p^{(2)}_{xz}+p^{(2)}_{ \bar x z}+
 p^{(5)} _{zx} + p^{(6)}_{\bar z x},
\\
q^{a,b} &=&p^{(7)}_{\bar z x} + p^{(8)}_{ z x},
\\
q^{a,c} &=& p^{(5)}_{\bar z x} + p^{(6)}_{z x},
\\
q^{b,b} &=& p^{(2)} _{x \bar z } +p^{(3)}_{xz}.
\label{eq2}
\end{eqnarray}
where $p_P+p_M$ in $q^a$ corresponds to the preparation and measurement errors
of the ancilla qubit for the syndrome measurement, and the subscripts 
$z$, $\bar z$ , $x$, and $\bar x$ mean $\sum _{A= Y,Z}$, $\sum _{A= I,X}$, $\sum _{A= X,Y}$,
and $\sum _{A= I,Z}$ respectively.
For example, $p^{(l)}_{z\bar x}= \sum _{A=Y,Z} \sum _{B=I,Z} p^{(l)}_{AB}= p^{(l)} _{YI}+p^{(l)} _{ZI}+p^{(l)} _{YZ}+p^{(l)} _{ZZ}$.

In the original scheme \cite{Raussendorf07a},
two-qubit gate errors $A\otimes B$ ($A,B = I,X,Y,Z$) and single-qubit gate errors $A$ ($A=X,Y,Z$) 
are assumed to occur with equal probability $p_g/15$ and $p_g/3$, respectively.
Thus $p^{(l)}_{AB}=p_g/15$ for all $A,B$ when $l=2,4,6,8$,
and $p^{(l)}_{IZ}=p^{(l)}_{ZX}=p^{(l)}_{ZY}=6p_g/15$ 
and $p^{(l)}_{AB}=p_g/15$ for other $A,B$ when $l=1,3,5,7$.
The preparation and measurement error probabilities are taken as $p_P=p_M=p_g$.
Then, Eqs. (\ref{eq1})-(\ref{eq2}) read
$q^{a} = 46p_g/15$,
$q^{b} = 44p_g/15 $,
$q^{c} = 44p_g/15$,
$q^{a,b}=q^{a,c}= q^{b,b}= 8p_g/15$.
On the other hand, the threshold value for $p_g$ is given by $p_g = 0.75\%$ in Ref. \cite{Raussendorf07a},
which is obtained by using the minimum-weight-perfect-matching algorithm.
This leads to threshold conditions of these probabilities (sufficient conditions for fault-tolerance) as
\begin{eqnarray}
q^a<0.023, \;\; q^b=q^c< 0.022, 
q^{a,b}=q^{a,c}=q^{b,b} <0.0040.
\nonumber \\
\label{eq3}
\end{eqnarray}

In the present architecture, on the other hand,
we use the TTGs of type I (for $l=1,5$), II (for $l=2,4,6,8$), and III (for $l=3,7$), where the state preparation for the syndrome measurement
and Hadamard operations are
incorporated (see TTG I and III).
Specifically, for the TTG of type I (i.e. $l=1,5$),
the error probabilities $p^{(l)}_{ab}$ ($a,b =z,\bar z, x ,\bar x$)
are given by
\begin{eqnarray}
p^{(l)}_{zx} &=& 4p_g/15+p_M,
\\
p^{(l)}_{z \bar x} &=& \bar F_2 + \bar F_3 + 4p_g/15,
\\
p^{(l)}_{\bar z x} &=& 4p_g/15,
\\
p^{(l)}_{xz} &=& p^{(l)}_{x\bar z}=0, 
\\
p^{(l)}_{\bar xz} &=& \bar F_1 + \bar F_2 + 8p_g/15,
\end{eqnarray}
where we take $p_{AB} = p_g/15$.
Similarly, for the TTG of type II and III (i.e. $l \neq 1,5$),
they are given by
\begin{eqnarray}
p^{(l)}_{zx} &=& p^{(l)}_{xz} = 4p_g/15+p_M ,
\\
p^{(l)}_{z \bar x} &=& \bar F_2 + \bar F_3 + 12p_g/15,
\\
p^{(l)}_{\bar z x} &=& p^{(l)}_{x\bar z}= 4p_g/15,
\\
p^{(l)}_{\bar xz} &=& \bar F_1 + \bar F_2 + 12p_g/15.
\end{eqnarray}
By using the these,
Eqs. (\ref{eq1})-(\ref{eq2}) and threshold conditions (\ref{eq3}) read
\begin{eqnarray}
q^a &=& 4(\bar F_2+\bar F_3)+\frac{40}{15}p_g +p_M < 0.023,
\\
q^b &=& 2(\bar F_1+ \bar F_2)+\frac{40}{15}p_g +2p_M < 0.022,
\\
q^c &=& 2(\bar F_1+\bar F_2)+\frac{32}{15}p_g +2p_M < 0.022,
\\
q^{a,b} &=&  q^{a,c} = q^{b,b} = \frac{8}{15}p_g +p_M<0.0040 .
\end{eqnarray}
(Note that $p_P=0$ in the present case, since preparation of $|+\rangle$ ancilla
for the syndrome measurement is taken into account in the TTG of type I.)
The threshold curve $(F,p_g=p_M)$ are plotted
in Fig. 3 (b), which is calculated with $\bar \mathbf{F}$
obtained by using the transition probability tensors.
Specifically, in the limit of $F \rightarrow 1$,
the output fidelity of the purification
is obtained as
$(\bar F_1, \bar F_2, \bar F_3)= (4p_g/15,2p_g/15,2p_g/15)$
in the leading order \cite{FY09}.
With $p_M = p_g$,
we obtain the following conditions on $p_g$:
\begin{eqnarray}
q^a &=& \frac{71}{15}p_g  < 0.023 \Leftrightarrow p_g < 0.0049,
\\
q^b&=& \frac{82}{15}p_g < 0.022 \Leftrightarrow p_g < 0.0040,
\\
q^c&=& \frac{74}{15}p_g  < 0.022  \Leftrightarrow p_g < 0.0045,
\\
q^{a,b} &=&  q^{a,c} = q^{b,b} = \frac{23}{15}p_g <0.0040 
 \Leftrightarrow p_g <0.0026.
 \nonumber \\ 
\end{eqnarray}
This leads to the physical threshold $p_g = 0.26 \%$ 
with $F \sim 1$.
Since $q^{a,b,c}$ are smaller than the threshold values for them,
the true physical threshold would be higher than $p_g = 0.26\%$,
which would be determined by using a full numerical simulation.

\section{Resource analysis}
The operational overhead under the physical error probability at 1/3 of the topological threshold is plotted as a function of the circuit size in Fig. 11 of Ref. \cite{Raussendorf07a}.
It reads that 
if we want to perform $\Omega = 3\times 10^{11}$ $\pi/8$ gates accurately
(i.e. an accuracy of $\sim \Omega ^{-1}$ is required for each $\pi/8$ gate),
a logical $\pi/8$ gate requires $T=2 \times 10^{10}$ ($O_3$ in Ref. \cite{Raussendorf07a}) physical two-qubit gates.

\begin{figure*}
\centering
\includegraphics[width=70mm]{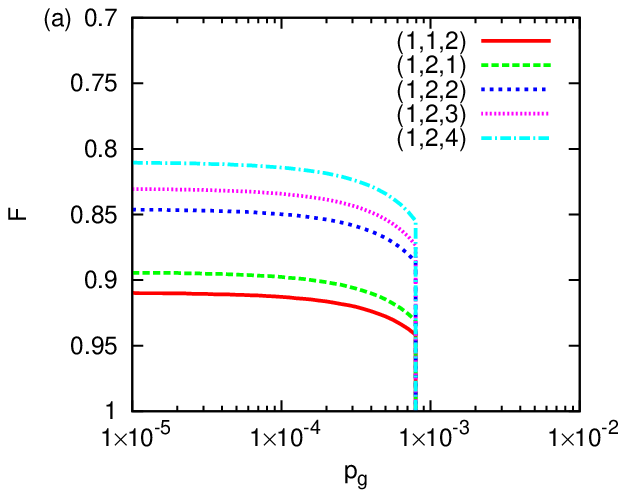}
\includegraphics[width=70mm]{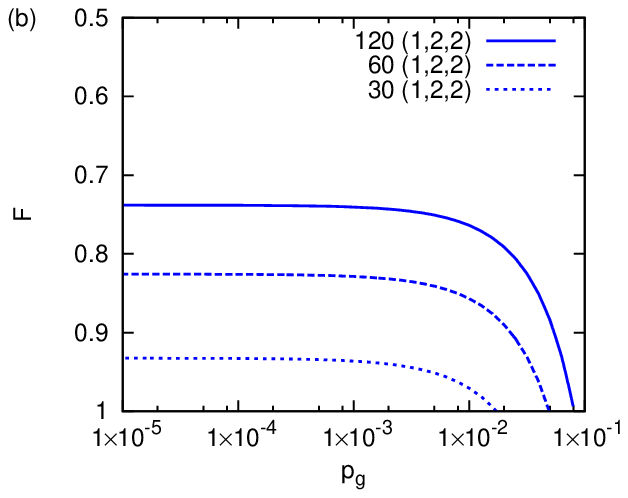}
\caption{$(F,p_g)$ at 1/3 of the topological threshold and operational overheads per TTG. (a) The channel fidelity $F$ and error probability $p_g=p_M$ of local operations
with which the independent and correlated error probabilities 
are at most 1/3 of the topological threshold are
plotted for each $(n_1,m_1,m_2)=(1,1,2),(1,2,1),(1,2,2),(1,2,3),(1,2,4)$.
(b) The contours of the total amount $K$ of local operations plus
quantum communication (i.e. the number of initial MESs) per TTG are 
plotted with respect to the channel fidelity $F$ and error probability $p_g=p_M$ of local operations for $(n_1,m_1,m_2)=(1,2,2)$.}
\label{resource}
\end{figure*}
In Fig. \ref{resource} (a),
1/3 of the topological thresholds in the present case
[i.e., the curves $(F,p_g)$ which satisfy
$q^a < 0.0049/3$, $q^b<0.0040/3$, $q^c<0.0045/3$ and 
$q^{a,b}=  q^{a,c} = q^{b,b}<0.040/3$]
are plotted for each $(n_1,m_1,m_2)=(1,1,2),(1,2,2),(1,2,3),(1,2,4)$.
It can be seen that $(n_1,m_1,m_2)=(1,2,2)$
achieves 1/3 of the topological threshold
even with $F \sim 0.9$ and $p_g \sim 0.1\%$.
The the total amount $K$ of local operations plus
quantum communication (i.e. the number of initial MESs) 
per TTG (i.e. per purified MES)
is determined for given repetition numbers $(n_1,m_1,m_2)$ 
by the channel fidelity $F$ and error probability $p_g=p_M$
of local operations
through the success probabilities $p_{\rm Lv1}$, $r_{\rm Lv1}$, and $r_{\rm Lv2}$.
In Fig. \ref{resource} (b),
the contours $K=30,60,120$ of the total amount $K$ for $(n_1,m_1,m_2)=(1,2,2)$
are plotted against $F$ and $p_g=p_M$. 
Specifically, $K \simeq 40$ when $F \sim 0.9$ and $p_g \sim 0.1\%$.
Accordingly, the total operational overhead 
can be calculated as $R=KT \sim 40 \times 2 \times 10^{10} \times 3 \times 10^{11} 
\sim  2 \times 10^{23}$.

\end{document}